\documentclass[prl,floatfix,twocolumn,showpacs,amsmath,amssymb]{revtex4}
\usepackage{graphicx}
\usepackage{dcolumn}
\usepackage{bm}
\begin{document}
\title{Tetramerisation of a frustrated spin-1/2 chain}

\author{Federico~Becca and Fr\'ed\'eric~Mila} 
\affiliation{
Institut de Physique Th\'eorique, Universit\'e de Lausanne,
CH-1015 Lausanne, Switzerland}

\author{Didier~Poilblanc}
\email{Didier.Poilblanc@irsamc.ups-tlse.fr}
\affiliation{Groupe de Physique Th\'eorique,
Laboratoire de Physique Quantique, UMR--CNRS 5626\\
Universit\'e Paul Sabatier, F-31062 Toulouse, France} 

\date{\today}

\begin{abstract}

We investigate a model of a frustrated
spin-1/2 Heisenberg chain coupled to adiabatic phonons
with a general form of magnetoelastic coupling. 
For large enough frustration and lattice coupling
a new tetramerised phase with three different bond lengths
is found. We argue that the zig-zag spin-1/2 chain LiV$_2$O$_5$
might be a good candidate to observe such a phase.

\end{abstract}

\maketitle

\begin{figure}
\vspace{2mm} 
\includegraphics[width=0.45\textwidth]{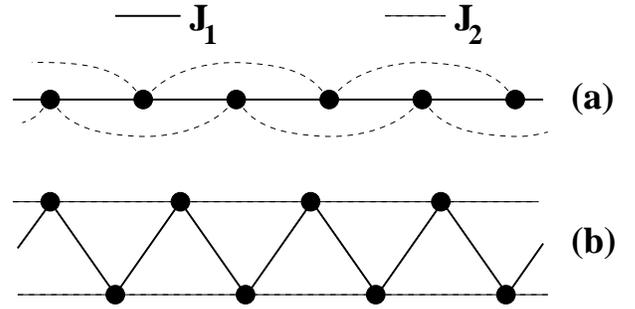}
\caption{\label{fig:lattices}
(a) Linear $J_1{-}J_2$ chain; (b) Zig-zag $J_1{-}J_2$ chain.
}
\end{figure}

Quasi one-dimensional (1D) quantum antiferromagnets exhibit 
fascinating magnetic properties at low temperatures. 
Inorganic compounds such as CuGeO$_3$ (Ref.\onlinecite{review}) 
or LiV$_2$O$_5$ (Ref.\onlinecite{LiV2O5})
are almost ideal prototypes of the spin-1/2 frustrated chain, the so-called
antiferromagnetic (AF) Heisenberg 
$J_1{-}J_2$ chain (see Fig.~\ref{fig:lattices}).
The chemistry of these compounds 
enables the synthesis of single crystals much larger than 
their organic analogs and 
consequently the achievement of new experimental studies.
Recently, the discovery of a
spin-Peierls (SP) transition in 
CuGeO$_3$~(Ref.\onlinecite{hase}) has drawn 
both experimental and theoretical interest. 

At temperatures larger than the interchain couplings 
the quasi-1D compounds CuGeO$_3$ or LiV$_2$O$_5$
are well described
as independent AF Heisenberg chains including next-nearest neighbor
(NNN) interactions responsible for frustration.
The nearest neighbor (NN) $J_1$ and NNN $J_2$ exchange integrals 
can be estimated by a fit of the
magnetic susceptibility, the high
temperature behavior being governed by $J_1$ and the position of the
maximum by the frustration ratio $J_2/J_1$. 
Values such as $J_1\approx 160$~K and 
$J_2/J_1\approx 0.36$ have been proposed for
CuGeO$_3$~(Ref.\onlinecite{rieradobry}).
On the other hand, in LiV$_2$O$_5$ the spin-1/2 V$^{4+}$ ions
form double-chains similar to Fig.~\ref{fig:lattices}(b) well separated by 
inert double-chains of V$^{5+}$ ions. Quantum chemistry
calculations suggest that $J_2$ could even be larger than $J_1$ in 
that case~\cite{DFT}.

The SP transition is an instability due to magnetoelastic effects
which is characterized (below a critical temperature $T_{\rm SP}$)
by the opening of a spin gap and the appearance of 
a lattice dimerisation. It was first predicted to occur in the
non-frustrated S=1/2 chain~\cite{SP_mechanism}, but 
the properties of the $J_1{-}J_2$
chain suggest that it is also a natural instability in that
case since the ground-state of that model is spontaneously
dimerized for $J_2/J_1 \gtrsim 0.24$. This is particularly clear at the 
so-called Majumdar-Ghosh point~\cite{MG} (MG) $J_2/J_1=0.5$, 
where the ground state (GS) is
two-fold degenerate, corresponding to two possible
dimerisation patterns formed by a succession
of disconnected singlet dimers.
However, when $J_2/J_1$ becomes very large, another instability
could occur: The $J_2$ chains are only weakly coupled, and they
could undergo a SP transtion of their own. The interplay between
both instabilities has not been considered so far.

In this Letter, we investigate on equal footings the role of
the frustration and of the lattice coupling.
Special emphasis is put on the search for new 
phases which would result from the combination of both 
effects. The competition between various orderings which could
eventually appear simultaneously can only be addressed 
by going beyond pertubative approaches.
Using Exact Diagonalisation techniques
we report evidences for 
a new mixed phase with both dimerisation and tetramerisation
amplitudes. 
Lastly, we discuss our results in the context
of the quasi-1D antiferromagnets CuGeO$_3$ and LiV$_2$O$_5$.

The Hamiltonian of a frustated spin chain on $L$ sites coupled to 
(adiabatic) lattice displacements is written as,
\begin{eqnarray}
\label{eq:ham} 
{\cal H}&=&\frac{1}{2}K \sum_{i} \delta_{i}^2
+ J_1\sum_{i} (1-A_1\delta_{i})\,
\vec{S}_{i} \cdot \vec{S}_{i+1} \nonumber \\
&+& J_2 \sum_i [1-A_2(\delta_{i}+\delta_{i+1})]  
\vec{S}_{i} \cdot \vec{S}_{i+2} \, ,
\end{eqnarray}
\noindent
where $\delta_i$ is the distortion of the bond between site
$i$ and $i+1$, $K$ the spring 
constant and the first term corresponds 
to the elastic energy loss. In general, this term might also contain
cross-terms such as $\delta_{i}\delta_{i+1}$ (depending on the
underlying geometry of the structure). We have checked 
that they do not affect the basis physics of this model
so that we omit them for simplicity.
Unless specified otherwise, $J_1$ sets the
energy scale. 
The spin-lattice couplings $A_a$ are assumed to be dimensionless 
so that the distortions $\delta_i$ are given 
in units of the lattice spacing.
Note also that, as can be seen from a trivial re-definition of the $\delta_i$,
the coupling strengths can be re-defined by the
reduced variables ${\tilde A}_1=A_1 (J_1/K)^{1/2}$ and
${\tilde A}_2=A_2(J_1/K)^{1/2}$ and used to investigate the phase diagram.
However, the ``physical'' values for the modulations $\delta_i$
depends on the $A_a$'s and $K$ separately.
Estimations of these parameters can be given on
physical grounds as will be discussed 
later on. Values such that $A_2=A_1$
and $A_2=2 A_1$ are relevant for the physical systems 
we are interested in.

\begin{figure}
\vspace{2mm} 
\includegraphics[width=0.32\textwidth,angle=-90]{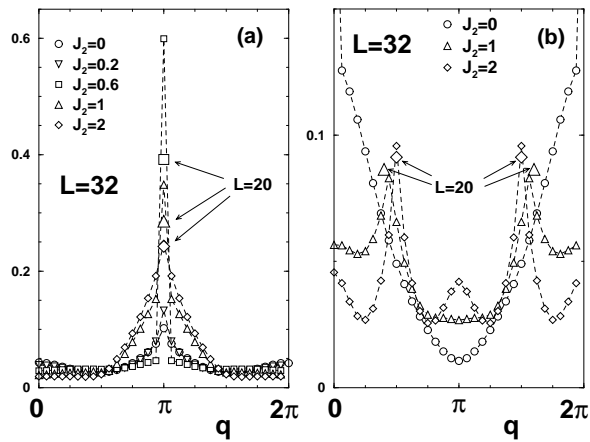}
\caption{\label{fig:suscep}
Generalised dimer susceptibilities as a function of chain momentum $q$
calculated by ED on a 32 site ring. 
Some peak amplitudes for $L=20$ are also shown as
indicated on the plot. (a) NN dimer susceptibility 
($a=1$); (b) NNN dimer susceptibility ($a=2$). 
}
\end{figure}

The phase diagram of the frustrated chain in the absence of 
lattice couplings ($A_1=A_2=0$) is well known.
The GS is uniform for small frustation (with power-law decay
of the spin correlations)
and becomes dimerized (with a finite spin gap) for
$J_2/J_1$ larger than a critical value~\cite{haldane}
$j_c$ which has been determined with great accuracy
by numerical methods,  
$j_c\simeq 0.241167$~(Ref.~\onlinecite{j1j2_num}). 
Interestingly enough, incommensurate spin correlations (away from 
the AF momentum $q=\pi$) appear for $J_2/J_1 > 0.5$ 
(Ref.~\onlinecite{incommensurate}).   

Before investigating the full Hamiltonian (\ref{eq:ham}), it is 
instructive to first consider ``generalised'' dimer susceptibilities 
of the form,
\begin{equation}
\label{eq:suscep} 
{\tilde S}(q,a)= \big< (\frac{1}{L} \sum_j 
\vec{S}_{j} \cdot \vec{S}_{j+a} \exp{(iqr_j)})^2 \big>_0,
\end{equation}
\noindent
where the expectation value $\big< ... \big>_0$ 
is taken in the GS of the $J_1{-}J_2$
chain in the absence of lattice coupling.
Physically, any instability towards a modulated dimer phase involving
dimers at distance $a$ would be signaled by a sharp peak of ${\tilde S}(q,a)$
at a given $q$ associated to the wavevector of the modulation.
As seen in Fig.~\ref{fig:suscep} (Ref.~\onlinecite{note1}), 
sharp peaks are indeed 
seen in ${\tilde S}(q,1)$ 
and ${\tilde S}(q,2)$ at momentum $q=\pi$ and $q=\pi/2$ respectively
signaling proximity of instabilities toward the formation of
dimerised ($q=\pi$) and tetramerized ($q=\pi/2$) phases involving
NN and NNN dimers respectively. Note that NNN $q=\pi/2$ dimer correlations
increase with increasing frustration $J_2/J_1$
while the maximum NN dimer
susceptibility occurs around $J_2/J_1 \simeq 0.5$.
Since the order parameter $\big<\frac{1}{L} \sum_j 
\vec{S}_{j} \cdot \vec{S}_{j+2} \exp{(i\frac{\pi}{2}r_j)}\big>$ of the
tetramerised phase is directly coupled to $A_2$
in Hamiltonian~(\ref{eq:ham}), the finite magnetoelastic $A_2$
coupling is then the key feature of the model.

In order to solve Hamiltonian~(\ref{eq:ham}) including the
magnetoelastic coupling we use Lanczos diagonalisations
of small finite rings of size $L$ with periodic boundary conditions.
An iterative procedure is used to determine 
the displacements $\delta_i$
by solving a set of coupled non-linear equations~\cite{iterative},
\begin{eqnarray}
\label{eq:non_linear} 
K \delta_{i} &-& J_1 A_1 \big< \vec{S}_{i} \cdot 
\vec{S}_{i+1} \big> \\
&-& J_2 A_2 
(\big< \vec{S}_{i} \cdot \vec{S}_{i+2} \big> 
+ \big< \vec{S}_{i-1} \cdot \vec{S}_{i+1}\big>)  =0, \nonumber
\end{eqnarray}
\noindent
where $\big< ... \big>$ is the expectation value in the
GS of Hamiltonian~(\ref{eq:ham}). Note that no translation
symmetry is {\it a priori} assumed in order to search for
lattice modulations of arbitrary periodicity (compatible with 
system size). We found that, generically,
the lowest energy lattice configuration
is obtained for a single or a superposition of 
the following distortions;
(i) a uniform (negative) component $\delta_i=\delta_0$ (which is due to the
finite compressibility of the system~\cite{note2}),
(ii) a dimerisation $\delta_i=\delta_D (-1)^i$ and (iii) a 
tetramerisation $\delta_i=\delta_T \cos{(\frac{\pi}{2}i+\phi_T)}$.
Although, the tetramerisation could be either site-centered 
(with $\phi_T=\pi/4$) or bond-centered (with $\phi_T=0$), only the second
bond-centered type (i.e. a modulation of the bonds 
like $\delta_T$--0--($-\delta_T$)--0) was found. 
This particular pattern can easily be understood
in the large-$J_2$ limit which consists of
two weakly coupled $J_2$ Heisenberg chains
(see Fig.~\ref{fig:lattices}(b)). In that limit, the magnetoelastic 
coupling $A_2$ tends to produce a dimerisation of
each chain so that $(\delta_{2p}+\delta_{2p+1})\propto (-1)^p$
which can indeed be realized by a tetramerisation of the chain
with $\phi_T=0$.

\begin{figure}
\vspace{2mm} 
\includegraphics[width=0.32\textwidth,angle=-90]{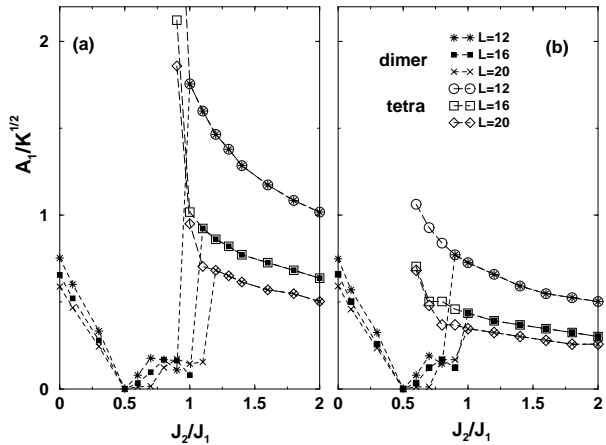}
\caption{\label{fig:critical_lines}
Phase diagram in the ${\tilde A}_1$ -- frustration plane for different
system sizes $L$. The data points (and the dotted lines)
correspond to the critical values of the (reduced) magnetoelastic
coupling above which dimerisation and/or tetramerisation (as shown
on the plot) appear; (a) $A_2/A_1=1$; (b) $A_2/A_1=2$.
In the latter case, a very small region of uniform 
phase inside the dimerised phase area (omitted here for clarity) was found
but shown to be a spurious finite size effect.
}
\end{figure}

\begin{figure}
\vspace{2mm} 
\includegraphics[width=0.45\textwidth,angle=0]{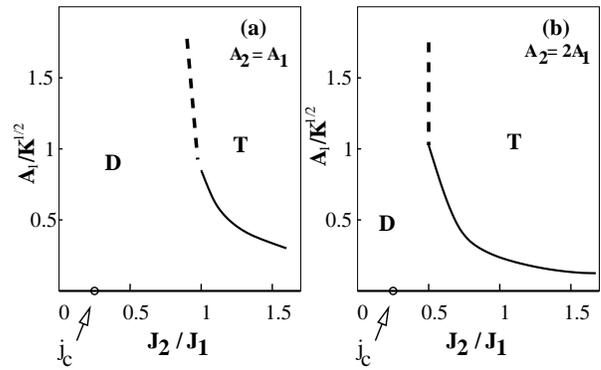}
\caption{\label{fig:phase_diag}
Tentative phase diagrams in the ${\tilde A_1}$ -- frustration plane 
obtained from a finite size scaling analysis 
of the data of Fig.~\ref{fig:critical_lines}. 
The dimerized and tetramerized phases 
are labelled by ${\bf D}$ and ${\bf T}$ respectively.
Thick dashed (full) lines correspond to first order-like (second order)
transition lines.
(a) $A_2/A_1=1$; (b) $A_2/A_1=2$. 
}
\end{figure}

The domains of stability of the various phases are shown in 
Fig.~\ref{fig:critical_lines} for two values of the parameter $A_2/A_1$.
Data are shown for cyclic rings of size $L=12$, $16$ and $20$ 
so that tentative phase diagrams
can be obtained (see Fig.~\ref{fig:phase_diag}) from a finite size 
scaling analysis. Various regimes have to be distinguished 
for these extrapolations. First, when $J_2/J_1<j_c$, 
the Heisenberg chain is critical and one expects
that a dimerised GS would be stabilized for arbitrary
magnetoelastic coupling~\cite{SP_mechanism}. 
Indeed, the finite critical value of
the coupling ${\tilde A}_1$ exhibits a clear $1/\sqrt{L}$ power-law behavior
with system size. In the range $j_c<J_2/J_1<0.5$
the system size dependence of the critical value 
of ${\tilde A}_1$ becomes exponential signaling the fact that 
the infinite Heisenberg chain forms singlet dimers,
even in the absence of
the lattice. Note that at the special MG point $J_2/J_1=0.5$ 
the critical coupling for ${\tilde A}_1$ vanishes for all sizes.
Special care is needed to analyse the data for $J_2/J_1>0.5$:
In a narrow range of $J_2/J_1$, $0.5<J_2/J_1<j_{\rm crit,1}$,
which depends on the $A_2/A_1$ ratio  
the system only dimerises above a small critical value of the coupling
$A_1$. In a range $j_{\rm crit,1}<J_2/J_1<j_{\rm crit,2}(L)$
(which extends with increasing 
system size), when ${\tilde A}_1$ exceeds a higher critical value, 
a tetramerisation superposes
to the existing dimerisation. 
For $J_2/J_1>j_{\rm crit,2}(L)$ dimerisation and tetramerisation
occur both for the same critical value of the coupling constant.
However, our data are consistent with the fact that 
$j_{\rm crit,2}(L)\rightarrow\infty$ when $L\rightarrow\infty$ so that
this last regime seems irrelevant.
For $0.5<J_2/J_1<j_{\rm crit,2}(L)$, the critical coupling for dimerisation
rapidly vanishes with increasingly large system sizes 
(although some increase has been observed for small sizes). 
For $J_2/J_1>j_{\rm crit,1}$ (e.g., $j_{\rm crit,1}\sim 0.85$
for $A_2=A_1$), the finite size dependence of the 
critical coupling for tetramerisation is consistent with a rapid
exponential behavior converging to a finite value.
An estimation of the infinite size
phase diagram is then possible with reasonable accuracy 
as shown in Fig.~\ref{fig:phase_diag}. In summary, our calculations predict 
that the $J_1{-}J_2$ chain is always dimerised once it is coupled
to the adiabatic lattice. Tetramerisation, however, occurs only for
large enough frustration and when the  lattice
couplings ${\tilde A}_1$ and ${\tilde A}_2$ exceeds some
critical values which vanish when $J_2/J_1$ increases to infinity.
Interestingly enough, our data also suggest that the
nature of the D $\rightarrow$ T transition might change when frustration 
increases, from first order (with discontinuities in
the dimerisation and tetramerisation amplitudes) to a continuous
second order-like line at large $J_2$.
Note that, even for large $J_2/J_1$ (limit of weakly
coupled chains), one still expects finite critical lattice couplings
since, in the absence of the lattice, 
an arbitray small perturbation $J_1$ introduces an
exponentially small gap and a finite spin correlation length.

We finish by discussing some applications of the present study to real
materials. The magnetoelastic couplings are generically 
due to strong dependence of the
exchange integrals with respect to distance, typically 
$J_a(r) \propto r^{-\alpha_a}$, with an exponent $\alpha_a$ in the range 7--15.
A small change of length $\delta{\vec r}$ of the bond connecting
two sites at distance $r_a$ along some direction ${\vec u}_a$ 
(${\vec r}=r_a{\vec u}_a$) leads to a linear change
of the AF coupling,
\begin{equation}
J_a(\delta{\vec r})=J_a(1-\frac{\alpha_a}{r_a}{\vec u}_a\cdot\delta{\vec r}).
\label{eq:first_order}
\end{equation}
\noindent
In the case of the linear chain  of Fig.~\ref{fig:lattices}(a) where 
the displacements occurs along the chain direction, Eq.~(\ref{eq:first_order})
predicts $A_2=A_1/2$ assuming the same values of $\alpha_a$ for the two
chemical bonds. In the case of CuGeO$_3$, the superexchange path 
giving rise to $J_2$ involves more intermediate states (in particular
Germanium orbitals) so that one expects $\alpha_2 > \alpha_1$ and 
$A_2\simeq A_1$ seems more physical in that case.
According to the phase diagram of Fig.~\ref{fig:phase_diag}(a), 
conditions for a small tetramerisation seem clearly not realised in CuGeO$_3$.
Indeed, for a frustration $J_2/J_1\sim 0.4$ and a small physical value
for the dimensionless coupling ${\tilde A}_1$, the $A_2$ coupling
becomes irrelevant (apart from producing a tiny overall contraction of the
lattice) and we expect a simple dimerisation 
$\delta_D\propto A_1J_1/K$.
Assuming $A_1\sim 10$, $J_1\sim 100\,{\rm Kelvins}$ and $K\sim 10\,{\rm eV}$
a dimerisation around $0.1$-$0.3\%$ of the lattice spacing is expected
in agreement with X-ray diffraction experiments~\cite{xrays}.

\begin{figure}
\vspace{2mm} 
\includegraphics[width=0.32\textwidth,angle=-90]{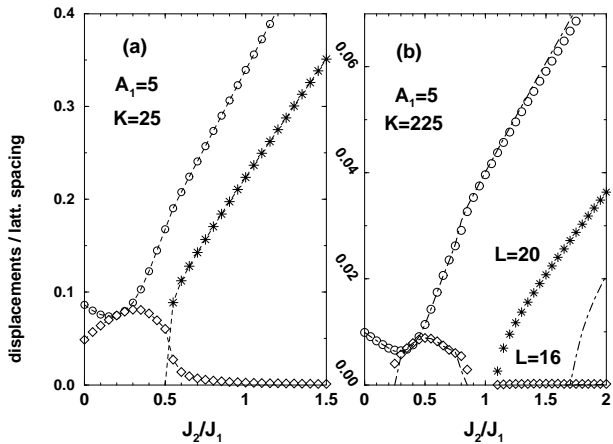}
\caption{\label{fig:magnitudes}
Amplitudes of the relative bond length change $|\delta_0|$ (open circles),
dimerisation $\delta_D$ (diamonds) and tetramerisation $\delta_T$ (stars)
(in units of the lattice spacing) versus frustration calculated on 
$L=20$ site rings for $A_1=5$ and $A_2=10$. Only non-zero
amplitudes are shown; (a) ${\tilde A}_1=1$; (b) 
${\tilde A}_1=1/3$: data for $L=16$ (thick dot-dashed lines)
are also shown in that case to indicate finite size effects. 
}
\end{figure}

We now turn to the case of the LiV$_2$O$_5$ compound.
If one assumes that atomic displacements 
in Fig.~\ref{fig:lattices}(b) would occur along the zig-zag chain direction
and that the exponents $\alpha_a$ are identical for the
two bonds, Eq.~(\ref{eq:first_order}) implies that
$A_2=2A_1$. As shown in Figs.~\ref{fig:phase_diag}(a-b), the 
stability of the tetramerised phase increases with increasing ratio
$A_2/A_1$. Hence, due to larger $A_2/A_1$ and $J_2/J_1$ ratios,
LiV$_2$O$_5$ seems, contrary to CuGeO$_3$, an interesting candidate 
for the new tetramerised phase. 
Typical exponents like $\alpha_1=\alpha_2=10$
gives $A_1=5$ and $A_2=10$. 
Assuming a physical value $J_1=400\,{\rm Kelvins}$ for the exchange constant 
and choosing 
$K=1 {\rm eV}$ 
($=10\, 000\,{\rm Kelvins}$) and $K=9\,{\rm eV}$ 
($=90\, 000\,{\rm Kelvins}$), 
we get 
${\tilde A}_1=1$ and ${\tilde A}_1=1/3$ respectively. 
The magnitudes of the dimerisation 
and tetramerisation for these parameters are shown in 
Fig.~\ref{fig:magnitudes} as a function of the frustration
$J_2/J_1$. For rather large magnetoelastic couplings such as the one used in 
Fig.~\ref{fig:magnitudes}(a) where finite size effects are negligible
we observe, for increasing magnetic frustration, a transition from a purely
dimerized phase to a new phase with a dominant tetramerisation and a
small dimerisation component.
For realistic couplings, let's say $K/J_1 > 200$, as 
seen in Fig.~\ref{fig:magnitudes}(b), finite size effects become large.
Nevertheless we expect a behavior similar to
that of Fig.~\ref{fig:magnitudes}(a) although 
with much smaller lattice displacements. Typically, while $\delta_T$ 
might be of the order of a percent of the lattice spacing,  $\delta_D$
is expected to remain much smaller.

To conclude, from numerical calculations
we have obtained the generic properties of the 
frustrated spin-1/2 chain coupled to adiabatic phonons.
Our results are confronted to experimental systems such as the
(quasi) linear CuGeO$_3$ chain and the zig-zag LiV$_2$O$_5$ chain.
While the observed small dimerisation of the SP phase of CuGeO$_3$ 
can easily be accounted for within such a simple model, we 
argue that LiV$_2$O$_5$ is a good candidate to observe 
a novel tetramerised phase with a doubling of the zig-zag chain 
periodicity (leading to new superstructure peaks in 
diffraction experiments) and two (slightly) non-equivalent V$^{4+}$ sites, 
a feature which could be observed in NMR-experiments. 

\acknowledgments
D.~P.  thanks the
{\it Institut de Physique Th\'eorique}
(Universit\'e de Lausanne) where part of this work was carried out
for hospitality 
and acknowledges support from the {\it Fondation Herbette} and the
Swiss National Fund.
F.~B. also thanks L.~Capriotti, A.~Parola, and S.~Sorella for 
helpful discussions.


\begin{thebibliography}{99}

\bibitem{review} For a review see e.g. J.P. Boucher and
    L.P. Regnault, J. Phys. I (Paris) {\bf 6}, 1939 (1996).

\bibitem{LiV2O5} M. Isobe and Y. Ueda, J. Phys. Soc. Jpn. {\bf 65}, 
3142 (1996); N. Fujiwara {\it et al.}, Phys. Rev. B {\bf 55}, R11945 (1997).

\bibitem{hase} M. Hase, I. Terasaki, and K. Uchinokura, 
     Phys. Rev. Lett. {\bf 70} 3651 (1993).

\bibitem{rieradobry} J. Riera and A. Dobry, Phys. Rev. B {\bf 51},
     16098 (1995); see also G. Castilla, S. Chakravarty, and V.J. Emery, Phys.
      Rev. Lett. {\bf 75}, 1823 (1995).

\bibitem{DFT} R. Valenti {\it et al.}, Phys. Rev. Lett. {\bf 86}, 5381 (2001).

\bibitem{SP_mechanism} M.C. Cross and D.S. Fisher, Phys. Rev. B 
{\bf 19}, 402 (1979); T. Nakano and H. Fukuyama, J. Phys. Soc. Jpn. {\bf 49},
1679 (1980).

\bibitem{MG} C.K. Majumdar and D.K. Ghosh, J. Math. Phys. {\bf 10}, 
1399 (1969).

\bibitem{haldane} F.D.M. Haldane, Phys. Rev. B {\bf 25}, 4925 (1982);
Phys. Rev. B {\bf 26}, 5257 (1982).

\bibitem{j1j2_num} S. Eggert, Phys. Rev. B {\bf 54}, 9612 (1996);
see also K. Okamoto and K. Nomura, Phys. Lett. A {\bf 169}, 433 (1992).

\bibitem{incommensurate} S.R. White and I. Affleck, Phys. Rev. B 
{\bf 54}, 9862 (1996).

\bibitem{note1} Note that, for simplicity,
only the $S_i^z S_{i+a}^z$ component have been included.

\bibitem{iterative} For details see 
A.E. Feiguin {\it et al.}, Phys. Rev. B {\bf 56}, 14 607 (1997); 
This method has also been applied to investigate
impurities in SP systems, see e.g.,
P. Hansen {\it et al.}, Phys. Rev. B {\bf 59}, 13 557 (1999).
We have also used a more polyvalent steepest-gradient method
and found identical results. 

\bibitem{note2}  A trivial 
renormalisation (increase) of the bare exchange couplings,
$\tilde J_1=J_1(1+A_1|\delta_0|)$ and $\tilde J_2=J_2(1+2A_2|\delta_0|)$
follows.

\bibitem{xrays} J.-P. Pouget {\it et al.}, Phys. Rev. Lett. {\bf 72},
4037 (1994).

\end{thebibliography}
\end{document}